\documentclass[useAMS]{mn2e}
\usepackage{psfig}

\usepackage{times}
\def\spose#1{\hbox to 0pt{#1\hss}}
\newcommand\lsim{\mathrel{\spose{\lower 3pt\hbox{$\mathchar"218$}}
     \raise 2.0pt\hbox{$\mathchar"13C$}}}
\newcommand\gsim{\mathrel{\spose{\lower 3pt\hbox{$\mathchar"218$}}
     \raise 2.0pt\hbox{$\mathchar"13E$}}}
\def\ltsima{$\; \buildrel < \over \sim \;$}
\def\lsim{\lower.5ex\hbox{\ltsima}}
\def\gtsima{$\; \buildrel > \over \sim \;$}
\def\gsim{\lower.5ex\hbox{\gtsima}}

\def\sch{Schwarzschild}

\title[Pairs in blazars and radio-galaxies]
{Electron positron pairs in blazar jets and $\gamma$--ray loud radio--galaxies}

\author[Ghisellini] 
{G. Ghisellini\thanks{E--mail:
gabriele.ghisellini@brera.inaf.it} 
\\
INAF -- Osservatorio Astronomico di Brera, via E. Bianchi 46, I--23807 Merate, Italy \\
}

\begin{document}


\pagerange{\pageref{firstpage}--\pageref{lastpage}} \pubyear{2007}

\maketitle

\label{firstpage}

\begin{abstract}
The matter content of extragalactic relativistic jets is still an unsolved issue.
There are strong arguments against pure electron--positron pair jets, 
but pairs could outnumber the electrons associated with protons by a factor 10--20.
This impacts on the estimate of the jet kinetic power, by reducing it by the 
same factor, and on the total energy delivered to leptons by the 
particle acceleration mechanism.
Pairs cannot be created in the same jet--zone responsible for
the high energy $\gamma$--ray emission we see in blazars, because 
the reprocessing of the created pairs would overproduce the X--ray flux.
Copious pair creation could occur in the inner zone of the
still accelerating jet, where the bulk Lorentz factor is small.
It is found that the inner zone can produce a sufficient number of pairs to 
replenish the zone of the jet where most of the luminosity is emitted, but only
if the $\gamma$--ray luminosity of the inner jet is above $10^{44}$ erg s$^{-1}$ 
at $\sim 1$ MeV.
Since the beaming is modest, this emission can be observed at large viewing angles,
and detected in radio--galaxies 
and lobe dominated quasars
at the flux level of $10^{-12}$--$10^{-11}$ erg cm$^{-2}$ s$^{-1}$
for a source  at a redshift $z=0.1$.
\end{abstract}

\begin{keywords}
galaxies: active--galaxies: jets--galaxies ---
radiation mechanisms: non--thermal --- gamma--rays: theory
\end{keywords}

\section{Introduction}

Relativistic extragalactic jets carry their power in 
the form of particles and fields.
Estimating how many leptons and protons
and the importance of the magnetic field flowing in the jet
is not trivial, since these  values depend on the
model we use to explain the radiation we see, and
to the presence of electron--positron pairs.
The latter quantity has been a subject of investigation for a long time,
with some consensus emerging about the scenario in which pairs are 
indeed present, but not dominating the dynamics of the jet.
This can be translated into a pair to proton number ratio of the order
of a few tens.
The arguments that have been put forward are the following.
\begin{enumerate}

\item
The spectra of powerful blazars are dominated by the high energy $\gamma$--ray
component:
the synchrotron luminosity is a factor 10--100 smaller.
This translates in a limit on the importance of the Poynting flux
(e.g. Celotti \& Ghisellini 2008), that cannot dominate the jet power
at the scale where the $\gamma$--ray emission is produced.
In turn, this implies that, at these scales, the jet must be matter
dominated. 

\item
If we have to produce enough pairs to contribute to the total leptonic component
in the jet region where most of the [100 MeV -- 100 GeV] $\gamma$--ray luminosity 
is emitted, the source should be opaque for
$\gamma$--$\gamma$ collisions in the entire [100 MeV -- 100 GeV] band.
This requires a large photon density, in turn implying a fast cooling 
for Compton scattering.  
Thus most of the absorbed luminosity is reprocessed to lower frequencies, 
especially in the X--ray band. 
In these cases the X--ray flux should be larger than observed
(Ghisellini \& Madau 1996).

\item
Therefore a large number of pairs must necessarily be produced 
in the inner jet (e.g. Blandford \& Levinson 1995; Sikora \& Madejski 2000).
In these regions radiative cooling is severe, and pairs
would quickly cool. 
If they are created copiously, such that their Thomson optical depth $\tau_+\gg 1$, 
they would quickly annihilate in a dynamical time
(i.e. the time needed for the region to double its size).
The surviving pairs would correspond to $\tau_+ \le 1$ in the inner jet
(Ghisellini et al. 1992).
This sets an upper limit to the number of pairs able to reach  
the parsec--scale, allowing Celotti \& Fabian (1992) to constrain the
amounts of pairs there.

\item
Cold leptons, traveling with a bulk Lorentz
factor $\Gamma$, would interact with the photons produced by the accretion
disk and re--isotropized by broad line clouds and/or scattered by 
free electrons external to the jet (Begelman \& Sikora 1987).
The resulting emission would be narrowly peaked, and mainly in the soft X--rays.
Observers within the $1/\Gamma$ beaming angle would see this radiation as
a X--ray ``bump" superimposed on the continuum (Sikora \& Madejski 2000;
Sikora et al. 1997; Moderski et al. 2004; Celotti, Ghisellini \& Fabian 2007).
The absence of this feature sets constraints on the total amount of leptons,
and argue against a jet dynamically dominated by e+e- pairs.

\item
If the main emission process of the 100 MeV -- 10 GeV radiation we detect from blazars
is the inverse Compton scattering of relativistic leptons off seed photons
produced externally to the jet, then the pattern of the radiation, as seen
in the comoving frame, is anisotropic (Dermer 1995).
More power is emitted in the upward direction, and the emitting region must recoil.
If the jet is ``light", only composed by e+e- pairs, the recoil would be enough
to stop the bulk motion of the jet.
In this way we can set an upper limit on the amount of pairs, that translates
in un upper limit of 10--20 on the pair to proton number ratio
(Ghisellini \& Tavecchio 2010).

\end{enumerate}

The presence of pairs in the region where most of the 
observed $\gamma$--rays are produced has important
effects on at least two issues.
The first is the jet power.
In powerful Flat Spectrum Radio Quasars (FSRQs),
if there is one cold proton per emitting lepton, then
the jet is dominated by the bulk kinetic power of protons, $P_{\rm p}$.
If instead there are $\xi \equiv (n_++n_{\rm p})/n_{\rm p}$ leptons per proton,
$P_{\rm p}$ can be reduced by the factor $\xi$. 
The second issue concerns the acceleration mechanism.
Shocks and reconnection processes are thought to accelerate
both leptons and protons. 
If the number of proton equals the number of leptons, 
at least half of the available energy goes to protons.
Instead, if pairs are present, most of the energy can go to leptons.

The total power budget of the jet can be reduced by the presence
of pairs, and the aim of the present study is to explore
the possibility to create pairs (through the $\gamma$--$\gamma \to e^\pm$ process)
in the inner zone of the jet,
where the plasma is still accelerating and the bulk
Lorentz factor $\Gamma$ is small.
The pairs that survive annihilation in this compact region can then 
refurnish the zone that produces the bulk of the radiation we see
from blazars, characterized by a larger $\Gamma$.
If this mechanism is common, there is a clear
observational consequence: misaligned sources, i.e. 
powerful radio--galaxies
and lobe dominated quasars,
should be strong emitters
in the hard X--ray to soft $\gamma$--ray region.
This is the diagnostic feature that provides
the observational test for this scenario.
The notation $Q=10^X Q_{X}$ (in cgs units) is adopted.


\section{The inner jet}

The most likely mechanism for the acceleration of extragalactic jets 
is through the conversion of an initially dominant Poynting flux into
kinetic power.
In the inner jet, at a few \sch\ radii, the jet is still heavily magnetically
dominated.
If some dissipation occurs, accelerating electrons to relativistic energies,
the resulting luminosity should be emitted by synchrotron radiation,
with severe radiative losses.
It is then likely that the emitting particles cannot reach high energies,
and their highest emitted synchrotron frequency is well below the pair
production threshold.
Therefore most of the high energy emission must be produced 
through the inverse Compton process.

In the inner jet the bulk Lorentz factor must be relatively small,
since the jet is just starting to accelerate.
This implies that the radiation produced externally to the jet (in the
disk, the corona, the broad line region) is not seen largely boosted
in the comoving frame (see Fig. 2 of Ghisellini \& Tavecchio 2009, GT09 hereafter).
In turn this implies that most of the radiation of the inner
jet is produced by the synchrotron process, unless it is self--absorbed.  
To produce a large amount of pairs we have these alternatives:
\begin{enumerate}
\item 
The synchrotron flux itself extends up to MeV energies.
This implies the presence of very energetic electrons,
which is problematic because radiative cooling, in these regions,
is very severe.
\item
Most of the synchrotron flux is optically thin, produced by electrons 
with energies reaching $\gamma\sim 100$.
One or two inverse Compton orders are enough to reach 1 MeV.
Since the region is magnetically dominated, we expect a large ratio
between the magnetic and radiation energy density, so that the synchrotron
self--Compton flux is at most comparable with the synchrotron luminosity.
\item
Most of the synchrotron flux is self absorbed, corresponding to small electron
energies.
In this case the high energy flux is produced by multiple scatterings.
Most of the luminosity cannot be emitted through the synchrotron
process, which is self--absorbed, and is emitted through the Comptonization process.
\item
Sikora \& Madejski (2000) considered yet another process to produce a significant luminosity
at $\sim$MeV energies, through the interaction of coronal X--ray photons with cold leptons
in the inner part of the jet (i.e. the ``protojet"). 
These interactions boost the original X--ray photons to MeV energies, i.e. above threshold.
The MeV photons are then absorbed and produce pairs; a fraction of these pairs 
can then load the protojet.

\end{enumerate}
Since there are several ways to generate the required MeV luminosity, 
it is convenient not to specify its origin.
We will therefore explore the observational consequences of the existence of
a MeV luminosity large enough to produce enough pairs to feed the jet
regions where most of the observed luminosity is produced.

We will then assume that in a conical jet of semi--aperture angle $\psi$,
a region located at a distance $R_0$ from the black hole 
produces an observed luminosity $L_0$, and that a fraction $f$ of it
is converted into electron positron pairs.
We will assume that the inner jet is accelerating according to
\begin{equation}
\Gamma_0\beta_0 \, =\, \max\left[\Gamma, \left( {R_0 \over 3 R_{\rm S} } \right)^{1/2} \right]
\label{gbeta}
\end{equation}
where $R_{\rm S}$ is the \sch\ radius and $\Gamma$ is the final bulk Lorentz factor.
The pair production process depends on the intrinsic compactness of the source, defined as
\begin{equation}
\ell^\prime_0 \, =\,  { \sigma_{\rm T} L_0 \over \psi R_0\delta_0^4 m_{\rm e} c^3 }  
\end{equation}
where $\delta_0 \equiv 1/[\Gamma_0 (1-\beta_0\cos\theta_{\rm v}]$ is the Doppler
relativistic factor at $R_0$ and $\theta_{\rm v}$ is the viewing angle.

The fraction $f$ depends on $\ell^\prime_0$ and on its spectrum.
For $\ell^\prime_0>60$ and for $\nu L_\nu$ spectra peaking in the MeV region,
$f$ reaches its maximum, which is nearly 0.1 (Svensson 1987).
If the spectrum peaks at energies much larger than 1 MeV most of the absorbed
energy is not transformed into rest mass of pairs, but in their kinetic energy, which is radiated away.
Therefore the conversion efficiency is less.
To maximize $f$, we will assume that the spectrum indeed peaks at $\sim$MeV energies,
but take into account the dependence of $f$ on the compactness.
Since the latter measures the optical depth for the pair production process, which 
becomes larger than unity for $\ell_0 \ge 60$, we simply have (Svensson 1987):
\begin{equation}
f = 0.1 \min\left[ 1, \, {\ell^\prime_0\over 60}\right] 
\label{f}
\end{equation}
The pair production rate $\dot n_+(R_0)$ is then
\begin{equation}
\dot n_+(R_0) \sim  {3 f L_0 \over 4\pi (\psi R_0)^3 \delta_0^4 m_{\rm e} c^2} 
\end{equation}
If the optical depth of the created pairs is $\tau_+<1$, annihilation is
negligible in one dynamical timescale (the time required for the region to expand 
doubling its radius) and we have
\begin{equation}
n_+(R_0) \sim \dot n_+(R_0) {\psi R_0\over \beta_0 c}, \qquad \tau_+(R_0)\le 1
\end{equation}
Instead, when $\tau_+\ge 1$, annihilation is faster than the dynamical timescale,
bringing the optical depth of the surviving pairs close to unity
(Ghisellini et al. 1992), giving
\begin{equation}
n_+(R_0) \sim  {1\over \sigma_{\rm T}\psi R_0}, \qquad \tau_+(R_0) > 1
\end{equation}
At $R_0$ a density $n_+(R_0)$ of pairs that survive annihilation is produced.
These pairs, traveling along the jet, do not suffer further annihilation, 
because their optical depth decreases.
Then they reach the region located at $R_{\rm diss} \gg R_0$, where
they are accelerated and produce a fraction of the flux we 
observed from a blazar. 
We want to estimate their contribution to the observed flux.

\section{Contribution of pairs to the blazar luminosity}

Assuming number flux conservation, the density of pairs reaching the 
region where most of the blazar luminosity is produced is
\begin{equation}
n_+(R_{\rm diss}) =  n_+(R_0)\, 
\left[ { R_0 \over  R_{\rm diss}} \right]^2 \, {\Gamma_0\beta_0 \over \Gamma\beta}
\label{nn}
\end{equation}
Let us assume that all these pairs, when reaching $R_{\rm diss}$, are accelerated
to relativistic energies, and form a particle density distribution $N_+(\gamma)$ 
as a result of continuous injection and radiative cooling.
The observed luminosity produced by pairs is:
\begin{equation}
L_{\rm +, obs} \, =\, V\delta^4 \int N_+(\gamma)\dot\gamma_{\rm c} m_{\rm e}c^2 d\gamma
\end{equation}
where $V$ is the volume of the emitting region, and $\delta$ is the 
corresponding Doppler factor.
If the emitting region is within the broad line region, located at 
a distance $R_{\rm BLR}$ from the black hole, the radiation energy 
density of the line photons is, as observed in the comoving frame:
\begin{equation}
U^\prime_{\rm ext}\, \sim \, \Gamma^2 { L_{\rm BLR} \over 4\pi R^2_{\rm BLR} c} 
\sim {\Gamma^2 \over 12 \pi}
\end{equation}
This assumes that $R_{\rm BLR} = 10^{17} L^{1/2}_{\rm d, 45}$ cm, and
that $L_{\rm BLR} = 0.1 L_{\rm d}$ (GT09; $L_{\rm d}$ is the disk luminosity).
Therefore, as long as $R_{\rm diss}<R_{\rm BLR}$, the external radiation energy 
density is constant and independent of $L_{\rm disk}$.
In powerful blazars this radiation energy density dominates over the 
energy density of internally produced synchrotron photons.
Thus the contribution of pairs to the $\gamma$--ray luminosity, due to Inverse 
Compton (IC), is:
\begin{equation}
\dot \gamma_{\rm c} \, =\, {4\over 3} {\sigma_{\rm T} c \gamma^2 U^\prime_{\rm ext} 
\over m_{\rm e} c^2}
\, =\, { \sigma_{\rm T} c\, \Gamma^2\over 9\pi m_{\rm e} c^2}\,  \gamma^2
\end{equation}
With these assumptions $L_{\rm +, \gamma, obs}$ becomes:
\begin{equation}
L_{\rm +,\gamma, obs} \, =\, {4 \sigma_{\rm T} c\, \over 27}\, (\psi R_{\rm diss})^3
\delta^4\Gamma^2 n_+(R_{\rm diss}) \langle \gamma^2\rangle 
\label{lgamma1}
\end{equation}
The factor $\langle\gamma^2\rangle$ depends on the particle distribution.
Since the observed high energy spectrum is peaked at a frequency $\nu_{\rm c}$,
the $N(\gamma)$ distribution must be a broken power law, with a break at
\begin{equation}
\gamma_{\rm b} \, =\, \left( {3 \nu_{\rm c} 
\over 4 \nu_{Ly\alpha} \Gamma\delta} \right)^{1/2}
\, \sim \, 300 \left( {h \nu_{\rm c} \over 100 \, {\rm MeV}}\right)^{1/2}
\left({1\over \delta_1\Gamma_1}\right)^{1/2}
\end{equation}
where we have assumed the $\nu_{Ly\alpha}$ as the typical frequency of the seed photons
(that are observed blueshifted by a factor $\sim \Gamma$ in the comoving frame).

\begin{figure}
\vskip -0.5cm
\hskip -0.5cm
\psfig{file=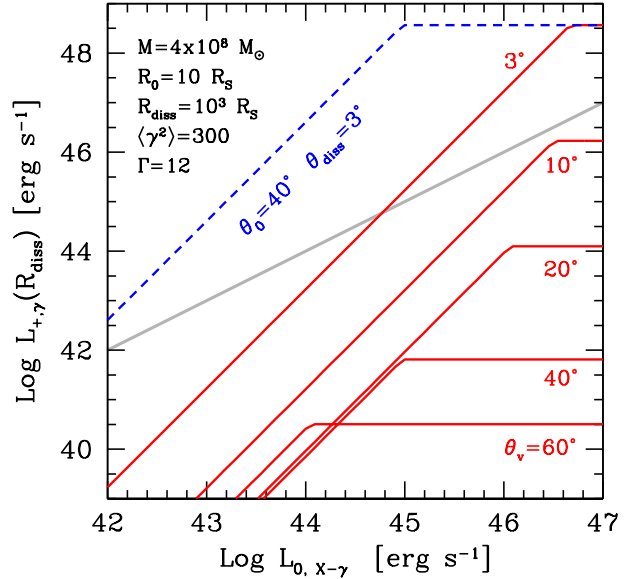,height=9cm,width=9cm}
\vskip -0.3 cm
\caption{Maximum $\gamma$--ray luminosity
produced at $R_{\rm diss}=10^3 R_{\rm S}$ by
the pairs that have been created at $R_0=10 R_{S}$ and
have survived annihilation there, according to Eq. \ref{lgamma1}.
At $R_0$, the source emits an observed luminosity 
$L_{0, x-\gamma}$ at high energies, and a fraction $f$ of
it is transformed into pairs (according to Eq. \ref{f}).
The bulk Lorentz factor at $R_{\rm diss}$ is $\Gamma=12$.
The different solid (red) lines are for different viewing angles $\theta_{\rm v}$,
as labeled.
In this case both the $R_0$ and the $R_{\rm diss}$ regions are observed at the same viewing angle, while for the dashed (blue) line the $R_{\rm diss}$ zone is assumed
to be oberved at $3^\circ$ (typical for a blazar), while the $R_0$ zone is observed at 
$40^\circ$ (typical for a 
{\bf
misaligned soure).}
The grey diagonal line indicates equality.
}
\label{test}
\end{figure}

The strong radiative cooling in powerful blazars implies that even low energy leptons
cool in one light crossing time.
This ensures that the low energy part of $N(\gamma)$ cannot be flatter than $\gamma^{-2}$.
And indeed the X--ray spectrum in these sources is often characterized by a power law
with a slope $F(\nu)\propto \nu^{-\alpha_x}$, with $\alpha_x\sim 0.5$, corresponding to
$N(\gamma)\propto \gamma^{-2}$.
As a rough estimate, we then have
\begin{equation}
\langle \gamma^2\rangle \, \sim \, {\int \gamma^2 N(\gamma)d\gamma \over \int  N(\gamma)d\gamma}
\approx \, \gamma_{\rm b} 
\end{equation}
Then the contribution of pairs to the observed $\gamma$--ray luminosity is
\begin{eqnarray}
L_{\rm +,\gamma, obs}  &=&  {4 \sigma_{\rm T} c\, \over 27}\, (\psi R_{\rm diss})^3
\delta^4\Gamma^2 n_+(R_{\rm diss}) \langle \gamma^2\rangle 
\nonumber \\
&\sim & {\sigma_{\rm T} \psi R_{\rm diss}   \gamma_{\rm b} \,    \over 9 \pi m_{\rm e} c^2}\,     
{\Gamma_0  \delta^4\Gamma  \over  \beta   \delta_0^4 }
  f L_0 , \qquad   \tau_+(R_0)\le 1
\nonumber \\
&\sim & 
{4   c\, \over 27}\, \psi R_{\rm diss}  \psi R_0    \, \gamma_{\rm b} \, 
{\Gamma_0\beta_0 \delta^4\Gamma  \over  \beta}, \,\, \,\, \tau_+(R_0) > 1
\label{lgamma2}
\end{eqnarray}
In principle, we can have three regimes:
i) at very low values of the intrinsic $L^\prime_0= L_0/\delta_0^4$, the
production of pairs is not ``saturated" (i.e. the $\tau_{\gamma\gamma}$ optical depth 
for pair production is less than unity, and $\ell_0^\prime<60$) 
and, furthermore, $\tau_+(R_0)<1$. 
In this case $L_{\rm +,\gamma, obs} \propto L_0^2$.
ii) If $\ell_0^\prime>60$, but still $\tau_+(R_0)<1$, then $L_{\rm +,\gamma, obs} \propto L_0$.
iii) If $\tau_+(R_0)>1$ then  $L_{\rm +,\gamma, obs}$ reaches its maximum
possible value  and is therefore independent of $L_0$.
In practice, case ii) corresponds to a very narrow range of luminosities, because
when $\ell_0^\prime$ becomes larger than 60, then the pair optical depth quickly becomes
larger than unity.
Fig. \ref{test} shows the amount of $\gamma$--ray luminosity produced 
by pairs at $R_{\rm diss}$, as a function of the $\gamma$--ray luminosity
produced by the inner ``protojet".
Different curves refer to different viewing angles.
For this illustrative example, the adopted parameters are:
the black hole mass $M=4\times 10^8 M_\odot$; $R_0=10 R_{\rm S}$,
$R_{\rm diss}=10^3 R_{\rm S}$, $\langle \gamma^2 \rangle=300$, 
and $\Gamma=12$.
The diagonal grey line indicates $L_{+,\gamma}(R_{\rm diss})= L_{0,X-\gamma}(R_0)$.
Both are observed luminosities.
The flat part of these curves corresponds to saturation in the number of pairs
that can survive annihilation (i.e. $\tau_+=1$).
As mentioned above, for typical parameters $L_{+,\gamma} \propto L^2_{0,X-\gamma}$.
For this choice of $\Gamma$, the $\sim$1 MeV luminosity of the ``protojet"
is larger than the luminosity produced at $R_{\rm diss}$
for viewing angles larger than $\sim 10^\circ$.

\begin{figure}
\vskip -0.8cm
\hskip -1.5cm
\psfig{file=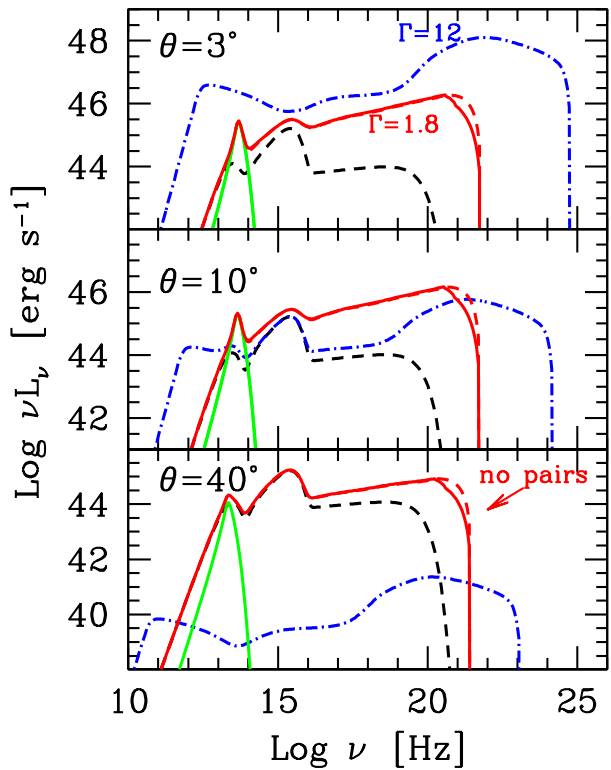,height=12cm,width=11.6cm}
\vskip -1 cm
\caption{Example of a SED produced in the inner portion of an
accelerating jet, at 10 \sch\ radii (solid red lines), where the bulk Lorentz
factor is $\Gamma= 1.8$,  and by a jet region located 
at $10^3 R_{\rm S}$ (dot--dashed blue line), where $\Gamma=12$.
The dashed (black) lines are the contribution by the accretion disk,
the IR torus and the X--ray corona.
The viewing angle is $\theta_{\rm v}=3^\circ$
(top panel); $\theta_{\rm v}=10^\circ$ (mid panel)
and $\theta_{\rm v}=40^\circ$ (bottom panel).
The intrinsic luminosity (i.e. as measured in the comoving frame)
is $L^\prime = 10^{44}$ erg s$^{-1}$ in the outer region, and 10 times that
for the inner component.
The inner region dissipates more power than the outer one,
but it is much less beamed.
The dashed (red) line is the spectrum before the absorption due to the
pair production process. 
The synchrotron emission of the inner jet is self absorbed (green solid line), 
and most of the radiation is produced by multiple inverse Compton scatterings.
For reference, $\nu L_\nu = 10^{44}$ erg s$^{-1}$ corresponds to
a flux of $4\times 10^{-12}$ erg cm$^{-2}$ s$^{-1}$ for a $z=0.1$ source,
while $\nu L_\nu = 10^{48}$ erg s$^{-1}$ corresponds to
a flux of $3.5\times 10^{-11}$ erg cm$^{-2}$ s$^{-1}$ for a $z=2$ source.
}
\label{inout}
\end{figure}

Consider a powerful blazar emitting a $\gamma$--ray luminosity $\sim10^{48}$
erg s$^{-1}$, observed at $\theta_{\rm v}\sim 3^\circ$.
If most of it is produced by pairs, the corresponding (observed) luminosity
produced in the inner jet must be slightly larger than $10^{46}$ erg s$^{-1}$.
If $\theta_{\rm v}\sim 3^\circ$  the beamed radiation 
produced at $R_{\rm diss}$ overwhelms the flux produced at $R_0$.
We can ask what a misaligned observer would see.
This is in part answered by the dashed line in Fig. \ref{test}: 
if the viewing angle if $\theta_{\rm v}=40^\circ$, we see a luminosity
from the inner jet close to $10^{45}$ erg s$^{-1}$.
Fig. \ref{inout} shows the SEDs corresponding to the two regions (located
at $R_0$ and $R_{\rm diss}$) for three different viewing angles.

The model used to construct these SEDs assumes that at $R_0$ the
jet is magnetically dominated, and that the leptons there have 
small (albeit relativistic) energies, and produce self--absorbed
synchrotron emission (solid green lines). 
Most of their energy is then released through multiple Compton 
scattering off these photons.
For the other features of the model, we refer the reader to GT09,
where they are fully described.
Following Eq. \ref{gbeta}, $\Gamma_0=1.8$ at $R_0$, and becomes 12
at $R_{\rm diss}$.
The dashed (red) lines at frequencies larger than $\sim$ 1 MeV indicate 
the spectrum before being absorbed by the $\gamma$--$\gamma \to e^\pm$ process.
The intrinsic (i.e. measured in the comoving frames) powers injected into the
regions are $10^{45}$ and $10^{44}$ erg s$^{-1}$ at $R_0$ and $R_{\rm diss}$,
respectively.
Due to beaming, the luminosity produced at $R_{\rm diss}$ is dominating
at all frequencies if the viewing angle is small, is comparable to 
the luminosity produced at $R_0$ for $\theta_{\rm v}\sim 10^\circ$,
and becomes unobservable for $\theta_{\rm v}=40^\circ$ or larger.
The pair production rate at $R_0$ is sufficient to yield $\tau_+>1$.
Therefore the surviving pairs correspond to an optical
depth of unity, thus to a density $n_+(R_0) = 1.2\times 10^{10}$ cm$^{-3}$.
Once arriving to $R_{\rm diss}$, the corresponding density (see Eq. \ref{nn}) is
$n_+(R_{\rm diss})\sim 10^5$ cm$^{-3}$.
If all these pairs are accelerated and emit, they would be enough to account
for the entire radiation produced at $R_{\rm diss}$.
However, we have already discussed that a pure pair plasma would face severe 
difficulties (the jet would stop due to Compton drag). 
Thus a small fraction of protons is likely to be present.

The other parameters used to construct these models are listed in Tab. \ref{para},
together with some derived quantities, such as the jet powers
in the form of radiation ($P_{\rm r})$, 
of Poynting flux ($P_{\rm B}$), relativistic emitting electrons ($P_{\rm e}$)
and cold protons ($P_{\rm p}$).
For the latter quantity we give two values: the first is assuming that there is
one proton per lepton, while for the second there is only one proton every 20 leptons.
The Poynting flux decreases from $R_0$ to $R_{\rm diss}$, in agreement with the
assumption that it is the cause for the jet acceleration.
As a consequence, the kinetic power $(P_{\rm p}+P_{\rm e})$ increases.

We alert the reader that what shown here is only an illustrative example.
There can be several factors affecting the exact values of the
used parameters.
As an example, both the emitting regions have been assumed to be homogeneous spheres,
with a tangled magnetic field that is taken to be representative of the jet magnetic field.
If the cause of the dissipation is magnetic reconnection, the listed $B$--field 
refers to the emitting region, and may be smaller than the magnetic field carried
by the jet.

\begin{table*} 
\centering
\begin{tabular}{llll llll llll lll}
\hline
\hline
Region     ~ &Size     &$P^\prime_{\rm i}$   &$B$ &$\Gamma$  
             &$\gamma_{\rm b}$ &$\gamma_{\rm max}$ &$s_1$  &$s_2$  &$\gamma_{\rm c}$ 
             &$\log P_{\rm r}$ &$\log P_{\rm B}$ &$\log P_{\rm e}$ &$\log P_{\rm p}$ &$\log(P_{\rm p}/20)$\\
~            &[1]   &[2] &[3] &[4] &[5] &[6] &[7] &[8] &[9] &[10] &[11]  &[12]   &[13] &[14]   \\
\hline   
$R_0$          &1.2 (10)     &1     &4949 &1.8 &7   &15   &--1  &2   &1  &45.42 &45.57 &43.30 &46.19 &44.89 \\   
$R_{\rm diss}$ &120 (1000)   &0.1   &2    &12  &100 &3e3  &--1  &2.5 &13 &46.11 &44.48 &45.39 &47.42 &46.12 \\
\hline
\hline 
\end{tabular}
\caption{
List of parameters for the SEDs shown in Fig. \ref{inout}.
For both models the accretion disk is assumed to emit a luminosity $L_{\rm d}=3\times 10^{45}$ erg s$^{-1}$
(at the 5\% Eddington level for a black hole mass of $M=4\times 10^8 M_\odot$).
The size of the BLR is $R_{\rm BLR}=1.7\times 10^{17}$ cm.
Col. [1]: dissipation radius in units of $10^{15}$ cm and (in parenthesis) in 
  $R_{\rm S}$ units;
Col. [2]: power injected in the blob calculated in the comoving frame, in units of $10^{45}$ erg s$^{-1}$; 
Col. [3]: magnetic field in Gauss;
Col. [4]: bulk Lorentz factor at $R_{\rm diss}$;
Col. [5] and [6]: break and maximum random Lorentz factors of the injected electrons;
Col. [7]: and [8]: slopes of the injected electron distribution [$Q(\gamma) \propto 
(\gamma/\gamma_{\rm b})^{-s_1} /[ 1+(\gamma/\gamma_{\rm b})^{-s_1+s_2} ]$;
Col. [9]: values of the minimum random Lorentz factor of those electrons cooling in one light crossing time;
Col. [10]--[13]: logarithm of the power in produced radiation ($P_{\rm r}$), 
Poynting flux ($P_{\rm B}$), leptons ($P_{\rm e}$), and cold protons ($P_{\rm p}$),
{\it assuming one proton per emitting electron;
} 
Col. [14] logarithm of the power of cold protons {\it assuming one proton per 20 emitting leptons.}
}
\label{para}
\end{table*}


%

\section{Powerful radio--galaxies and lobe dominated quasars}

Fig. \ref{inout} shows that if the inner jet is the producer of the pairs
that feed the $R_{\rm diss}$ regions of the jet,
then its luminosity must be rather large and not strongly beamed.
This statement is rather general {\it and independent of the specific adopted model.}

As a consequence, misaligned blazars should be 
very strong hard X--ray sources (if their inner jet produce pairs copiously).
If this scenario is correct, then 
powerful FSRQs, observed to emit $L_\gamma\sim 10^{48}$ erg s$^{-1}$, should have
misaligned counterparts emitting at a level of $10^{45}$ erg s$^{-1}$ around 1 MeV.
This means a flux of $\sim 4\times 10^{-11}$ erg cm$^{-2}$ s$^{-1}$ for 
a source at $z=0.1$.
This is not far from the sensitivities of current hard X--ray instruments, such
as the Burst Alert Telescope (BAT) onboard the {\it Swift} satellite (Ajello et al. 2009; 2012), 
and  reachable (albeit at lower energies) in even short 
exposures by the {\it Nu--Star} satellite (Harrison et al. 2005).
As a consequence, this scenario can be tested: if the inner jet regions
produce pairs to feed the outer dissipation jet region, 
then misaligned radio loud AGNs
should be strong emitters in the very hard X--ray range.

The luminosity of this $\sim$ MeV emission should be related to the
$\gamma$--ray luminosity produced at $R_{\rm diss}$,
but the latter is visible  
only 
in aligned objects, not in radio--galaxies
and lobe--dominated quasars.
Conversely, in aligned object we do not see the radiation produced
at $R_0$.
We need an isotropic component visible in both classes of objects,
that is related to the jet power, hence to the $\gamma$--ray luminosity
of blazars.
There can be two options.
The first is to compare energy of the radio lobes, assumed to be a calorimeter.
The difficulty here is to estimate the lobe lifetime, to infer the average
power of the jet.
A simpler opportunity could be the measurement of the mass accretion rate,
believed to be related to the jet power.
The latter could be measured 
by the broad emission lines. 
Their luminosity  gives
a good estimate of the entire disk emission, hence of the mass accretion rate,
hence of the jet power, hence of the $\gamma$--ray luminosity of an aligned 
source (i.e. a blazars, see e.g. Ghisellini et al. 2010; Sbarrato et al. 2012).

In other words: suppose that a radio--galaxy 
(or a lobe--dominated quasar)
is found to emit $10^{44}$ erg s$^{-1}$
at $\sim$1 MeV.
Suppose to see broad lines in its optical spectrum, enabling to estimate
a broad line region luminosity of $10^{43}$ erg s$^{-1}$.
The accretion disk should then be a factor 10 more luminous, so 
$L_{\rm d}\sim 10^{44}$ erg s$^{-1}$.
This is, very approximately, of the order of the power $P_{\rm r}\sim L_{\rm bol}/\Gamma^2$  
carried  by the jet in the form of radiation (Celotti \& Ghisellini 2008; GT09), where
$L_{\rm bol}$ is the observed jet bolometric luminosity.
If the latter is dominated by $L_\gamma$, 
we expect $L_\gamma\sim \Gamma^2 P_{\rm r} \sim \Gamma^2 L_{\rm d} 
\sim 10^{46}L_{\rm d, 44}\Gamma_1^2$ erg s$^{-1}$
(with an uncertainty of several).
We then conclude (see Fig. \ref{test}) that indeed the inner jet of the 
misaligned AGN
could produce enough pairs to sustain the expected
$\gamma$--ray luminosity of its aligned counterpart.
With these (admittedly very rough) arguments, we can relate the luminosity of
the inner jets of radio--galaxies 
and lobe--dominated quasars
with the $\gamma$--ray luminosities that would be seen by an aligned observer.

\section{Conclusions}

Electron--positron pairs cannot be produced in the same region where
most of the $\gamma$--ray luminosity of blazars is emitted,
but there are no strong arguments against their creation in the
inner part of the jet (at $\sim 10 R_{\rm S}$), where the bulk Lorentz factor is small.
The pairs that survive annihilation can feed larger regions of the jet
(the ``$\gamma$--ray zone" at $\sim10^3 R_{\rm S}$), and be re--accelerated
there, to contribute significantly to the emission we see from blazars.
The most efficient way to create pairs in the inner jet is through photon--photon collisions,
requiring a relatively large luminosity peaked around 1 MeV.
Since in these regions the bulk Lorentz factor is small, and the corresponding beaming
angle is large, these large luminosities are observable in 
nearby misaligned sources, i.e. radio--galaxies.
This is the observational test of the scenario: if a substantial number
of pairs are contributing to the luminosity of powerful blazars, than powerful
radio--galaxies should be strong hard X--ray emitters.

\section*{Acknowledgements}
I thank F. Tavecchio for useful discussions.

\end{document}